\documentclass[conference]{IEEEtran}
\IEEEoverridecommandlockouts

\newcommand{\hide}[1]{}

\usepackage{graphicx}
\usepackage{pgfplots}
\usepackage{pgf-pie}

\begin{document}

\title{Use and Perceptions of Multi-Monitor Workstations:\\ A Natural Experiment%
\thanks{This research was supported by the ISRAEL SCIENCE FOUNDATION (grant no.\ 832/18).}}


\author{\IEEEauthorblockN{Guy Amir$^*$ ~~~~ Ayala Prusak$^*$ ~~~~ Tal Reiss$^*$ ~~~~ Nir Zabari$^*$ ~~~~ Dror G. Feitelson
\thanks{$^*$Equal contribution.}}

\IEEEauthorblockA{Department of Computer Science\\
The Hebrew University of Jerusalem, 91904 Jerusalem, Israel}}

\maketitle

\begin{abstract}
    Using multiple monitors is commonly thought to improve productivity, but this is hard to check experimentally.
    We use a survey, taken by 101 practitioners of which 80\% have coded professionally for at least 2 years, to assess subjective perspectives based on experience.
    To improve validity, we compare situations in which developers naturally use different setups---the difference between working at home or at the office, and how things changed when developers were forced to work from home due to the Covid-19 pandemic.
    The results indicate that using multiple monitors is indeed perceived as beneficial and desirable.
    19\% of the respondents reported adding a monitor to their home setup in response to the Covid-19 situation.
    At the same time, the single most influential factor cited as affecting productivity was not the physical setup but interactions with co-workers---both reduced productivity due to lack of connections available at work, and improved productivity due to reduced interruptions from co-workers.
    A central implication of our work is that empirical research on software development should be conducted in settings similar to those actually used by practitioners, and in particular using workstations configured with multiple monitors.
\end{abstract}


\section{Introduction}

The use of multiple computer monitors has become ubiquitous in software development work environments.
This has been driven by the combination of affordable and decreasing prices, and the production of flat monitors, which require less desk space.
But do they indeed provide tangible benefits?

Using multiple monitors is part of a progression of practices aimed at improving workflows.
An important step was the ability to place jobs in the background, and switch between them.
Another was to define virtual workspaces, each complete with several windows, and switch between those \cite{henderson86}.
One use of such a setup is to allocate separate locations for tasks or subtasks \cite{ringel03}.
With multiple monitors the workspace locations no longer need to be virtual \cite{grudin01}.

Research about software engineering practices seems to have largely ignored the physical setup used by software developers.
A huge number of papers are published every year suggesting new tools for some specific software engineering task.
But there is little concern with how these tools mesh together to form a software development environment \cite{bulajic13}.
In practice, most software is developed in Integrated Development Environments (IDEs) such as Visual Studio, Eclipse, etc.,
and there has been some research on how such environments are used \cite{amann16,damevski17,murphy06}.
One important attribute appears to be the ability to integrate third-party tools, leading to the creations of an ecosystem of tools for each IDE.

The physical setups used to work with these environments---including computers, screens, furniture, and even office space---are another matter.
While ergonomics and office space arrangement are important topics in Business Management and Labor studies, there has been very little academic research on these issues in the context of software development \cite{mishra12}.
Perhaps the best known discussion in the one in DeMarco and Lister's \textit{Peopleware}, which emphasizes parameters like space and quietness \cite{demarco:people}.

But for the individual developer, the configuration of his or her personal development station is of the utmost importance.
Discussions on how to arrange the desktop occasionally appear in
unofficial
online forums,
including advice on arrangements, prideful YouTube videos and photos complete with background lighting, and praise for especially noteworthy setups.
As noted above, today many of these setups include multiple displays or ultrawide displays, including curved ones often favored by gamers.

Intuition suggests that multiple monitors should benefit productivity.
The added screen space can enable developers to have more information accessible and in sight, including documentation, code samples, design documents, interactions with co-workers, and more.
But do developers indeed find multiple monitors profitable?

Answering such questions experimentally is impractical.
Using multiple monitors is a basic element of one's work habits.
Requiring experimental subjects to switch from one setup to another within a limited-time experiment may therefore lead to threats to validity due to using unfamiliar working conditions.
Consequently we resort to exploiting natural experiments, where circumstances cause people to use different setups---e.g.\ when a difference exists between a developer's work and home setups.
Such situations have become all the more important recently due to the Covid-19 pandemic, which has sent many software developers to work from home.

Based on a survey of 101 practitioners, we collect data on work practices and on their perspectives on their work environment.
This included configuration preferences and adjustments made to customize their homes for work.
The results indicate that using multiple monitors is largely standard at work, and considered desirable also for home setups.
However, it is not the most important factor affecting productivity.

\section{Related Work}

Research that has been done on the use of multiple monitors falls into two main categories.
The first tries to assess the effect of using multi-monitor workstations on performance.

Czerwinski et al.\ conducted a user study with 15 participants to compare the use of a 15" screen with an experimental curved 42" wide display when performing a predefined sequence of office tasks \cite{czerwinski03}.
Despite the huge difference in monitor sizes, the results were that the time to perform all the tasks was reduced by a relatively modest 9\% when using the larger display.
But the subjective opinion of nearly all participants greatly favored the larger display.
These results foreshadow other studies including ours: people like larger displays, but it is hard to quantify concrete performance gains.

Colvin et al.\ conducted a much larger study with 108 participants, again using simulated office work as the task \cite{colvin04}.
The results again favored using multiple screens, with 6--16\% faster work and 33\% fewer errors.
But another important result was that the benefits depend on experience with the task.
This casts a shadow over experiments using predefined simulated tasks, which may not correspond to participants' typical work.

Kang and Stasko conducted a similar experiment, asking 28 university participants to complete a trip planning task using various office productivity tools \cite{kang08}.
Interestingly, the tasks also included interruptions: in one case, to handle an email message, and in another to respond to 2 instant messages.
Again, the results showed some benefit for using multiple monitors, and they were also favored by the participants.

Stegman et al.\ conducted an experiment with 36 participants performing office and engineering tasks \cite{stegman11,ling17}.
While the focus of the study was on using and managing multiple application windows, they also measured the time to perform the tasks.
Unlike previous experiments no significant difference was found between the times to complete the tasks using single or dual monitors.
However, with dual monitors there were fewer mouse clicks and window switches.

The second category of studies includes papers about the design of workspaces.
Henderson and Card studied the dynamics of using windows, and concluded that window usage exhibits locality of reference \cite{henderson86}.
This motivated their design of multiple distinct workspaces in the Rooms window management system.
But each individual task was limited to one ``room'', so this design was not compatible with exploiting multiple monitors.

Hutchings et al.\ were worried that the expanded screen space might increase the overhead devoted to window management, making the use of multiple monitors counterproductive and frustrating \cite{hutchings04}.
They therefore studied the dynamics of switching between widows and occlusion relations among windows.
One of their main results was the identification of a distinction between monitor uses, where one is used for interaction and the other mainly for display (e.g.\ an open email where you can see if something arrives) or background activity (e.g.\ playing music).

Grudin used a different approach, and conducted hour-long interviews with 18 users of multiple monitors \cite{grudin01}.
Notably, half of his subjects were involved in software development.
One of his observations was that additional monitors are used for peripheral activities, allowing a combination of retained focus on the primary monitor and noting other things on the secondary monitor but without suffering an interruption.
He also noted that even when developers use multiple monitors themselves, they do not design applications to support the effective use of multiple monitors.

Bi and Balakrishnan compared the use of single or dual monitors not to each other, but to the use of a huge 16'$\times$6' high resolution (6144$\times$2014 pixels) display covering an entire wall (created using an array of 6$\times$3 projectors) \cite{bi09}.
Moreover, the 8 participants used this setting for 5 consecutive days for their routine work (as opposed to the typical 1-hour sessions of artificial tasks in the experiments discussed above).
To preserve privacy, data was collected only on window and mouse events.
Participants were also asked to maintain an activity log, and note advantages and disadvantages of working with the large display.
The results were that all participants overwhelmingly preferred the large display.
However, participants who conventionally used a single-monitor setting preferred the large display 81\% of the time, whereas those who conventionally used dual monitors preferred it only 61\% of the time.
Interestingly, while dual-monitor users usually used one of their monitors as the focal point and the other as peripheral, when using the large display all participants used the center of the display as the focal point and the areas above and to both sides as peripheral.
Finally, all participants performed many more moving and resizing operations on windows when using the large display, but dual-monitor users also perform many more such operations than single-monitor users.

On a deeper level, Harrison and Dourish note that screen \emph{space} is merely an opportunity.
The real question is how to make it into a useful and meaningful \emph{place} for work \cite{harrison96}.
This can be used to theorize about how people organize their work environment, and in particular their desktops or monitors.
An important issue is the freedom to arrange things according to personal taste---just like the allocation of functions to rooms and the artefacts inhabiting these rooms are what makes a house into a home.
Multiple monitors may also afford a distinction between a place for individual work and a place for interaction with others.
Alternatively, this distinction may be temporal rather than spatial.

\section{Research Questions}

Many software companies today set up their development staff with multi-monitor workstations.
As developers, we can say from our personal experience that using a dual-monitor workstation ``feels better'' and is more satisfactory.
One reason is that multiple-monitors enable less mouse clicking.
A common scenario is when working on a project requiring frequent switching from the editor to other resources such as documentation, papers, etc.
Other reasons are better window management and a larger monitor area. 

We wanted to see whether these feelings are common among developers, and whether monitor configuration is believed to have an effect on productivity.
However, productivity is hard to pin down because it involves not only the quantity of code produced, but also its quality, in combination with the difficulty of the problems being solved.
It is also difficult to measure because it is affected by myriad factors \cite{wagner19}, and because developers have different working habits.
Experiments that require them to work in diverse settings---and in particular including settings to which they are not accustomed---are therefore unreliable.

For these reasons we decided to focus our study on analysing user experience,
and investigating the subjective reasons for preferring a certain monitor configuration.
To compare the use of single and multiple monitors we used a dual natural experiment: comparing developers who actually use single or multiple monitors in their routine work, and seeking developers who changed their setup due to the impact of the Covid-19 pandemic.

Our work is framed by the following research questions: 
\begin{itemize}
    \item \textbf{Physical configuration:}
    How common is it among developers to use multiple monitors?
    What is the preferred number of monitors used?
    Are there differences between work and home setups?
    \item \textbf{Monitor usage and productivity:}
    Does the use of multiple monitors affect perceived productivity?
    Does the usage of one monitor differ from the other?
    \item \textbf{Covid-19:}
    How did the pandemic affect programmers' work environment?
    What are the difficulties and benefits of working from home, compared to working from the office?
\end{itemize}

\section{Methodology}

Since software development productivity is difficult to measure in an objective way, we chose to focus on the subjective user experience of programmers working with different monitor configurations.
This was done using a survey, and investigating the preferences and experiences of developers regarding their monitor configuration.

\begin{figure}\centering
    \includegraphics[scale=0.5]{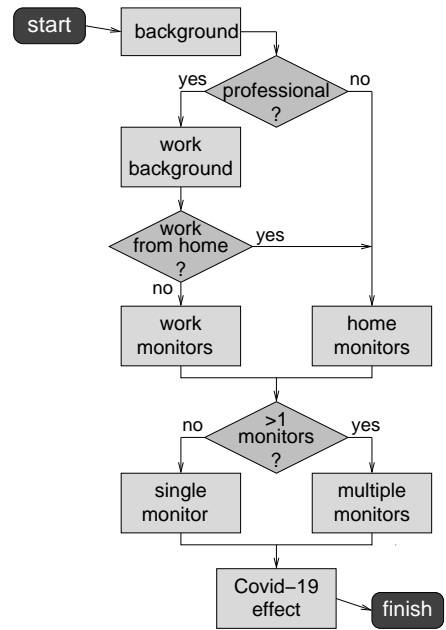}
    \caption{\sl Survey flow.}
    \label{fig:flow}
\end{figure}

In order for the research to be as broad as possible, we approached programmers in different professional positions and companies, in addition to computer science and engineering students.
We also kept the survey concise so as to keep subjects interested and truthful, and to make it appealing for users to answer.
Completing the survey typically took up to 10 minutes.

The survey included questions regarding the work environment during the Covid-19 pandemic and its effects on productivity compared to their usual work environment.
Many development teams nowadays are required to work from home, which may have caused developers to adjust to new work environments.
This also might have affected their work productivity.
Constructing a new home work space or adjusting an existing one may expose a developer's real preferences.
In our survey we gathered data about the developers' perspective regarding their new circumstances. 

The survey was written, filled, and analyzed using the Google Forms platform.
This platform was adequate due to our survey's simplicity and relatively small number of questions.

Our survey was aimed for diverse participants: people who code for a living and students, people who use multiple monitors and people who use just one.
We wanted to customize questions for each type of participant, to enable a deeper understanding of their considerations and perspectives.
In order to do this we constructed the survey to have a unique flow for each type of participants.
In other words, the answers given by each participant may lead to different blocks of questions, each targeted to different circumstances.
This enabled us to ask general questions regarding the monitor configuration of the participants, but also to ask questions specifically matching the participant's situation. 

The survey's structure (see Fig.\ \ref{fig:flow}) included an introduction block in which participants are divided into professional and non professional programmers.
Professional programmers are asked about their professional background, to enable an analysis of correlations between professional skills and workstations configurations.
This block included questions regarding the participant's position (back-end developer, web designer, etc.), rank in the development team (team leader or developer), and the company profile.
Then, participants are divided into those who usually work from home and those who usually work from the office.
Each group is asked about their work environment, and specifically their monitor use.
The last split is between single monitor and multiple monitor users.
This block contains questions about the subjective experience of productivity.
Finally all participants were asked about the Covid-19 pandemic and its effects.

\section{The Participants}

Our survey was filled by 101 participants.
Following Falessi et al.\ \cite{falessi18} we do not classify them according to their formal status as students or professionals, as this is an inaccurate oversimplification.
Instead, we asked them to report whether most of their coding was done at work, for studies, or as a hobby.
The result was that nearly $\frac{3}{4}$ of them code for a living, and most of the rest are investing in studies.

\hide{
\begin{figure}[h!]\centering
\begin{tikzpicture}
\pie[radius=1.4, color={red!70,blue!45,yellow!40,black!30},
     sum=auto, text=pin]
    {24/\textsf{studies},
     72/\textsf{work},
     2/\textsf{hobby},
     3/\textsf{other}}
\end{tikzpicture}
    \caption{Coding reasons of survey participants.
    Numbers are out of 101 responses.}
    \label{fig:participants}
\end{figure}
}

We also asked the participants for how many years they have coded professionally.
As shown in Fig.\ \ref{fig:vetek}, only 4 did not code professionally at all.
This indicates that the vast majority of those who are currently studying also have work experience.
But most of the participants have only up to 5 years experience.

\begin{figure}[h!]\centering
\begin{tikzpicture}
\pie[radius=1.2, color={black!30,red!70,orange!70,cyan!50,blue!70},
     sum=auto, text=pin]
    {4/\textsf{none},
     17/\textsf{1},
     55/\textsf{2-5},
     19/\textsf{5-10},
     6/\textsf{$>$10}}
\end{tikzpicture}
    \caption{Total years of professional coding of survey participants.}
    \label{fig:vetek}
\end{figure}

 Focusing on the participants who currently code professionally, they were predominantly from large companies: only 20 reported working for a company with 1--50 employees, whereas 22 came from companies with 100-500 employees, and an additional 26 from companies with more than 500 employees.

In terms of positions, they were quite varied.
13 each self-identified as full-stack developers and as involved in algorithms and machine learning.
12 identified as back-end developers (databases),
11 as application developers,
and 6 as research.
A few were system engineers, architects, embedded, or front-end.
None identified as QA (testers).
In terms of seniority, just over half of the survey participants self-identified as juniors.
A quarter identified as seniors, and half as many as technical or team leads.

\hide{
\begin{figure}[h!]\centering
\begin{tikzpicture}
\pie[radius=1.2, color={red!80,red!40,blue!50,violet!80,yellow!50},
     sum=auto, text=pin]
    {2/\textsf{student},
     38/\textsf{junior},
     19/\textsf{senior},
     10/\textsf{tech/team lead},
     2/\textsf{freelancer}}
\end{tikzpicture}
    \caption{Current position of working survey participants.
    Numbers are out of 71 responses.}
    \label{fig:position}
\end{figure}
}

The distribution of main coding languages was quite similar to the industry most popular languages:
nearly all participants listed python as one of their main coding languages, followed by C/C++, Java, and Javascript.

While we naturally can't claim our participants represent all developers, we believe the above distributions indicates that we have a reasonable sample that is not biased by the demands of a certain position or language.

\section{Results}

\subsection{Home Setup vs.\ Work Setup}

One of the background questions for participants who currently code professionally was about their work environment.
The usual work environment of the vast majority (87.5\%) of working participants was some form of shared space (Fig.\ \ref{fig:where}).
At the same time, a non-negligible minority (8.3\%) were already working from home, even before Covid-19.

\begin{figure}[h!]\centering
\begin{tikzpicture}
\pie[radius=1.2, color={blue!70,red!55,yellow!60,black!30},
     sum=auto, text=pin]
    {6/\textsf{home},
     18/\textsf{open space},
     45/\textsf{shared office},
     3/\textsf{other}}
\end{tikzpicture}
    \caption{Work environment of professional coding survey participants.
    `Other' includes private office.}
    \label{fig:where}
\end{figure}

Several questions considered differences between work and home setups, and the degree of satisfaction with these setups.
When asking participants about their desk size, participants working from the office tended to be more satisfied (Fig.\ \ref{fig:desk}).

\begin{figure}[h!]\centering
\begin{tikzpicture}
  \begin{axis}[ybar,
    width=5.5cm, height=2.5cm, scale only axis=true,
    enlarge x limits=0.15,
    ymin=0, ymax=60, ytick={0,20,40,60}, ylabel=\% of group,
    xtick=data, xtick pos=bottom, 
    legend pos=north west,]
	\addplot coordinates 
	{(1,3.1) (2,3.1) (3,7.8) (4,29.7) (5,56.3)};
	\addplot coordinates 
	{(1,5.4) (2,16.2) (3,18.9) (4,21.6) (5,37.8)};
	\legend{at work,at home}
  \end{axis}
  \node[yshift=-4mm] {low};
  \node[xshift=55mm, yshift=-4mm] {high};
\end{tikzpicture}
    \caption{Satisfaction with desk size.}
    \label{fig:desk}
\end{figure}

Asking about the number of monitors used, we can see most people working at the office use two monitors, while people working from home mostly use one (Fig.\ \ref{fig:work-home-mon}).
None of the participants used more than 2 monitors at home.

\begin{figure}[h!]\centering
\parbox{\linewidth}{\hspace*{15mm} home \hspace*{25mm} work \hspace*{8mm} monitors}
\begin{tikzpicture}
\pie[radius=1.0, color={red!80,blue!50},
     sum=auto, text=]
    {27/\textsf{1},
     10/\textsf{2}};
\pie[radius=1.0, pos={3.4,0}, color={red!80,blue!50,yellow!50},
     sum=auto, text=legend]
    {8/\textsf{1},
     48/\textsf{2},
     8/\textsf{3}};
\end{tikzpicture}
    \caption{Number of monitors used by survey participants for their main coding activity.
    Numbers are out of a total of 101 responses.}
    \label{fig:work-home-mon}
\end{figure}

Almost half (45.9\%) of the participants working from home believed that the number of monitors they use is smaller than what most use, while most (85.9\%) people working from the office believe the number of monitors they use is similar to most. This can be attributed to the fact that most people working from the office use what seems to be the standard number of two monitors, and also that when working from the office people are exposed to their peers' desk configuration and can reliably compare to their own.

\begin{figure}[h!]\centering
\begin{tikzpicture}
  \begin{axis}[
    xbar, xmin=0,
    width=7.0cm, height=4.9cm, 
    enlarge y limits=0.1,
    xlabel={number of participants},
    symbolic y coords={no home setup,same setup,other,more productive,need lab equip.,don't want more,cost,no space},
    ytick={no home setup,same setup,other,more productive,need lab equip.,don't want more,cost,no space},
    nodes near coords, nodes near coords align={horizontal},
    bar shift=0,
    bar width=8pt,
    ytick pos=left,
    ]
    \addplot coordinates {
        (22,no space)
        (19,cost)
        (7,don't want more)
        (2,need lab equip.)
        (2,more productive)
        (1,other) };
    \addplot coordinates {
        (27,same setup)
        (10,no home setup) };
  \end{axis}
\end{tikzpicture}
    \caption{Reasons cited for difference between work and home setup.
    Numbers from 71 working participants.}
    \label{fig:why-diff}
\end{figure}

Fig.\ \ref{fig:why-diff} shows the distribution of answers regarding possible differences between home and work setups.
The most commonly cited reasons were lack of space and high cost.
Being satisfied with the setup despite the difference came in a distant third.

In an open question about the considerations leading to the chosen setup, the most common answers regarding the work setup were that it was the default, and that it was convenient to do development on one screen and ``all the rest'' (Internet, search, documentation, communications, and more) on the other.
Actually using two different computers was also mentioned.
Regarding the home setup several additional considerations were mentioned, including
not needing multiple monitors because it is easy to move between desktops;
using a high-resolution retina screen makes others look bad;
a preference for using hand-written notes and sketches on the desk to actually design complicated stuff;
and a preference to use a laptop on the sofa.

\subsection{Multiple Monitors vs.\ Single Monitor}

As mentioned above, 75\% of participants used to working from the office use two monitors (12.5\% use one and the other 12.5\% use three), but only 27\% of participants working from home (all the rest use only one).
For comparison, Czerwinski et al.\ reported in 2003 that only up to 20\% of users used multiple monitors \cite{czerwinski03}.
For 66\% of those using multiple monitors these monitors are of the same size.

When asking participants if they would want to have more monitors, 63\% of those with one monitor said they would.
Among those who already had multiple monitors, 23\% said they want more.
The rest found their current setup to be optimal.
This strengthens our understanding that using two monitors is some sort of standard.
Within the limits of our sample, this standard largely seems to hold also when looking at different professional positions---with the possiblel exception of architects (Fig.\ \ref{fig:mon-who}).

\begin{figure}[h!]\centering
\begin{tikzpicture}
  \begin{axis}[
    xbar, xmin=0, xmax=3,
    width=7cm, height=4.5cm, 
    enlarge y limits=0.1,
    xlabel={average number of monitors},
    symbolic y coords={applications,full stack,algorithms/ML,back-end,research,syst. engineer,architect},
    bar width=8pt,
    ytick=data,
    ytick pos=left,
    ]
    \addplot coordinates {
        (1.8,applications)
        (1.82,full stack)
        (1.92,algorithms/ML)
        (2.09,back-end)
        (2.2,research)
        (2.25,syst. engineer)
        (2.67,architect) };
  \end{axis}
\end{tikzpicture}
    \caption{Average monitors used at work for different 
    positions. Only positions with at least 4 participants are shown.}
    \label{fig:mon-who}
\end{figure}

\subsection{Multiple Monitors Use}

\begin{figure}[h!]\centering
\begin{tikzpicture}
  \begin{axis}[ybar,
    width=6cm, height=3.3cm, 
    enlarge x limits=0.3,
    ymin=0, ymax=30,
    ylabel={participants},
    xtick=data, xtick pos=bottom, 
    title=Use one monitor more than others,
    title style={xshift=-18mm}]
	\addplot coordinates 
	{(1,7) (2,15) (3,16) (4,26) (5,2)};
  \end{axis}
  \node[yshift=-4mm] {all same};
  \node[xshift=45mm, yshift=-4mm] {only one};
\end{tikzpicture}

\begin{tikzpicture}
  \begin{axis}[ybar,
    width=6cm, height=3.3cm, 
    enlarge x limits=0.3,
    ymin=0, ymax=30,
    ylabel={participants},
    xtick=data, xtick pos=bottom, 
    title=Assign different tasks to each monitor,
    title style={xshift=-15mm}]
	\addplot coordinates 
	{(1,8) (2,8) (3,6) (4,15) (5,28)};
  \end{axis}
  \node[yshift=-4mm] {disagree};
  \node[xshift=45mm, yshift=-4mm] {agree};
\end{tikzpicture}

\begin{tikzpicture}
  \begin{axis}[ybar,
    width=6cm, height=2.8cm, 
    enlarge x limits=0.3,
    ymin=0, ymax=20,
    ylabel={participants},
    xtick=data, xtick pos=bottom, 
    ytick={0, 10, 20},
    title=Move windows between monitors a lot,
    title style={xshift=-15mm}]
	\addplot coordinates 
	{(1,6) (2,19) (3,16) (4,14) (5,11)};
  \end{axis}
  \node[yshift=-4mm] {disagree};
  \node[xshift=45mm, yshift=-4mm] {agree};
\end{tikzpicture}

    \caption{Usage of multiple monitors.}
    \label{fig:multi-use}
\end{figure}
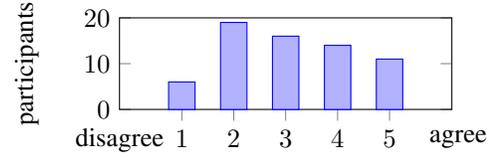

When asking participants who work with multiple monitors about their work habits, it appeared that usually some system is employed.
In other words, the use of multiple monitors is grounded in practical reasons, and is not just a common custom.
As Fig.\ \ref{fig:multi-use} shows, participants tend to use one monitor more than the other.
Moreover, they typically assign specific tasks to the different monitors, and tend not to move windows from one monitor to the other.
These results agree with work practices found in previous research \cite{bi09,grudin01,hutchings04}.

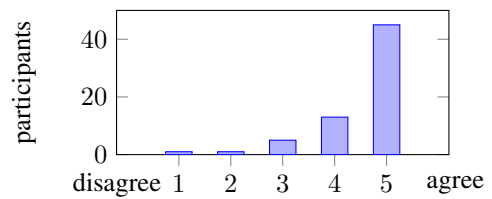
\begin{figure}[h!]\centering
\begin{tikzpicture}
  \begin{axis}[ybar,
    width=6cm, height=3.5cm, 
    enlarge x limits=0.3,
    ymin=0, ymax=50,
    ylabel={participants},
    xtick=data, xtick pos=bottom, 
    title=More productive with more than one monitor,
    title style={xshift=-13mm}]
	\addplot coordinates 
	{(1,1) (2,1) (3,5) (4,13) (5,45)};
  \end{axis}
  \node[yshift=-4mm] {disagree};
  \node[xshift=45mm, yshift=-4mm] {agree};
\end{tikzpicture}
    \caption{Perception of productivity with multiple monitors.}
    \label{fig:prod-multi}
\end{figure}

\begin{figure}[h!]\centering
\begin{tikzpicture}
\pie[radius=1.2, color={blue!70,yellow!50,red!60,black!30},
     sum=auto, text=pin]
    {1/\textsf{improved},
     10/\textsf{no change},
     52/\textsf{reduced},
     3/\textsf{other}}
\end{tikzpicture}
    \caption{Expectations regarding productivity change if forced to use only 1 monitor.}
    \label{fig:prod-to-1}
\end{figure}
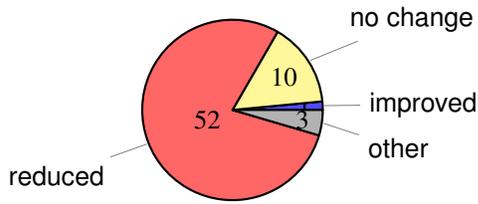

When asked about their subjective opinion on productivity when using multiple monitors, the vast majority of participants who do in fact use them believe it makes them more productive (Fig.\ \ref{fig:prod-multi}).
Conversely, when asked to imagine a scenario in which they would be required to shift to working with a single monitor, 78.8\% believed that this would harm their productivity (Fig.\ \ref{fig:prod-to-1}).

\begin{figure}[h!]\centering
\begin{tikzpicture}
  \begin{axis}[
    width=6.5cm, height=4cm, 
    ymin=0, ymax=100, 
    ylabel={\% in group},
    xlabel={years of professional coding},
    symbolic x coords={1,2-5,5-10,$>$10},
    xtick={1,2-5,5-10,$>$10},
    ]
    \addplot coordinates {
        (1,82)
        (2-5,69)
        (5-10,93)
        ($>$10,100) };
  \end{axis}
\end{tikzpicture}
    \caption{Percentage of participants thinking more monitors improve productivity, as a function of experience.}
    \label{fig:more-improve}
\end{figure}
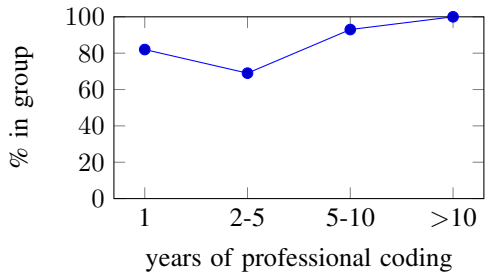

Interestingly, there seems to be some correlation between experience and the belief that multiple monitors enhance productivity.
Nearly all participants with 5 or more years of experience thought this is the case (Fig.\ \ref{fig:more-improve}).
Participants with up to 5 years of experience were less sure.

\section{Covid-19 Pandemic Effects}

As mentioned above, we feel that conducting a controlled experiment in which subjects are asked to perform a certain task using an unfamiliar setup would not produce reliable results.
But the global Covid-19 pandemic has created a natural experiment in which we can see what developers actually did when forced to change their working conditions.
We used this opportunity to ask about programmers' work environments and settings during the pandemic.

\begin{figure}[h!]\centering
\begin{tikzpicture}
  \begin{axis}[
    xbar, xmin=0,
    width=7.5cm, height=4.0cm, 
    enlarge y limits=0.2,
    xlabel={number of participants},
    symbolic y coords={no changes,other,leave of absence,changed hours,work from home},
    ytick={no changes,other,leave of absence,changed hours,work from home},
    nodes near coords, nodes near coords align={horizontal},
    bar width=8pt,
    bar shift=0,
    ytick pos=left,
    ]
    \addplot coordinates {
        (83,work from home)
        (40,changed hours)
        (5,leave of absence)
        (2,other) };
    \addplot coordinates {
        (11,no changes) };
  \end{axis}
\end{tikzpicture}
    \caption{Changes in work as a result of the Covid-19 pandemic. Numbers out of 101 total respondents.}
    \label{fig:covid-effect}
\end{figure}
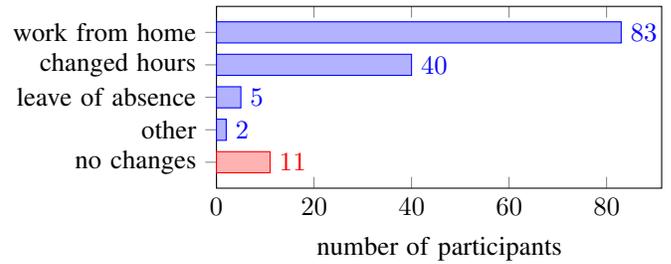

As shown in Fig.\ \ref{fig:covid-effect}, the main effect on the participants of our survey was that $\frac{5}{6}$ of them switched to working from home instead of at the office.
A large fraction also changed their working hours.
Only a few were placed on leave of absence, but this could be due to a bias in our sample, which emphasized large rich companies.

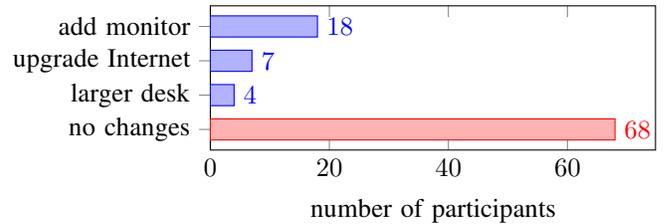
\begin{figure}[h!]\centering
\begin{tikzpicture}
  \begin{axis}[
    xbar, xmin=0,
    width=7.5cm, height=3.5cm, 
    enlarge y limits=0.2,
    xlabel={number of participants},
    symbolic y coords={no changes,larger desk,upgrade Internet,add monitor},
    ytick={no changes,larger desk,upgrade Internet,add monitor},
    nodes near coords, nodes near coords align={horizontal},
    bar width=8pt,
    bar shift=0,
    ytick pos=left,
    ]
    \addplot coordinates {
        (18,add monitor)
        (7,upgrade Internet)
        (4,larger desk) };
    \addplot coordinates {
        (68,no changes) };
  \end{axis}
\end{tikzpicture}
    \caption{Changes in home setup as a result of the Covid-19 pandemic. Numbers from 95 responses.}
    \label{fig:covid-home}
\end{figure}

The need to work (more) from home caused some of the participants to upgrade their home work environment (Fig.\ \ref{fig:covid-home}).
The most common change was to add a monitor.
This strengthens our assessment that multiple-monitor usage is perceived as beneficial by programmers.
However, the majority of participants did not make any such changes.

When asking about the pandemic's effects on productivity,
the largest group, containing 39.6\% or the participants, believed they were unaffected.
The second largest group, with 33.7\%, believed that their productivity had been impaired.
And a considerable minority of 26.7\% believed that their productivity had actually improved.

\begin{figure}[h!]\centering
\begin{tikzpicture}
\begin{axis}[
title=productivity change,
xbar stacked,
xmin=0,
width=7cm, height=3cm, 
enlarge y limits=0.4,
enlarge x limits=false,
bar width=15pt,
legend style={at={(0.5,-0.7)},
anchor=north,legend columns=-1},
xlabel={\% of participants in category},
symbolic y coords={students,professionals},
ytick={students,professionals},
ytick pos=left,
]
\addplot+ [xbar, draw=red, fill=red!60] coordinates {(41.4,students) (31,professionals)};
\addplot+ [xbar, draw=yellow, fill=yellow!50] coordinates {(37.9,students) (39.4,professionals)};
\addplot+ [xbar, draw=blue, fill=blue!70] coordinates {(20.7,students) (29.6,professionals)};
\legend{impaired, unchanged, improved}
\end{axis}
\end{tikzpicture}
    \caption{Productivity change due to Covid-19 for students vs.\ professionals.}
    \label{fig:covid-change}
\end{figure}
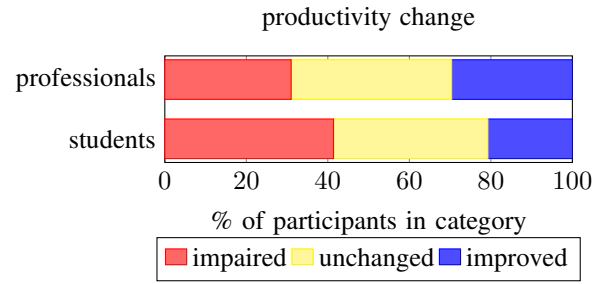

Interestingly, when examining the answers of students and professional programmers separately, we can see that the students were the more pessimistic:
Among students those who thought their productivity was harmed outnumbered those that thought it had improved by a ratio of 2:1, while among professionals the two groups were essentially equal (Fig.\ \ref{fig:covid-change}).

What might be the most interesting results regarding the effect of working from home due to Covid-19 are the factors that affect workers' productivity.
We expected the home environment to suffer from more interruptions and to provide inferior working conditions.
While these factors indeed exist, they were not the most prominent.

The main factor that was perceived as causing changes in productivity was interactions with co-workers at work.
But incredibly, this worked both ways.
74.2\% of those who said that working from home \emph{decreased} their productivity named the lack of connections to interfaces available just at work (e.g.\ co-workers) as one of the factors.
As the same time, 77.8\% of those who said that working from home \emph{increased} their productivity mentioned less interruptions from co-workers as a reason.
In other words, co-workers both help productivity and hinder it.

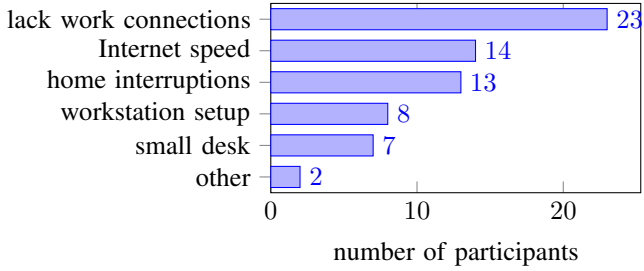
\begin{figure}[h!]\centering
\begin{tikzpicture}
  \begin{axis}[
    xbar, xmin=0,
    width=6.5cm, height=4.1cm, 
    enlarge y limits=0.1,
    xlabel={number of participants},
    bar width=8pt,
    symbolic y coords={other,small desk,workstation setup,home interruptions,Internet speed,lack work connections},
    ytick={other,small desk,workstation setup,home interruptions,Internet speed,lack work connections},
    nodes near coords, nodes near coords align={horizontal},
    ytick pos=left,
    ]
    \addplot coordinates {
        (23,lack work connections)
        (14,Internet speed)
        (13,home interruptions)
        (8,workstation setup)
        (7,small desk)
        (2,other) };
  \end{axis}
\end{tikzpicture}
    \caption{Reasons cited for reduced productivity when working from home, from 31 respondents.}
    \label{fig:home-bad}
\end{figure}

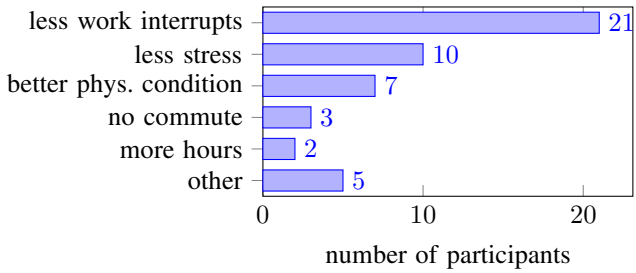
\begin{figure}[h!]\centering
\begin{tikzpicture}
  \begin{axis}[
    xbar, xmin=0,
    width=6.5cm, height=4.1cm, 
    enlarge y limits=0.1,
    xlabel={number of participants},
    bar width=8pt,
    symbolic y coords={other,more hours,no commute,better phys. condition,less stress,less work interrupts},
    ytick={other,more hours,no commute,better phys. condition,less stress,less work interrupts},
    nodes near coords, nodes near coords align={horizontal},
    ytick pos=left,
    ]
    \addplot coordinates {
        (21,less work interrupts)
        (10,less stress)
        (7,better phys. condition)
        (3,no commute)
        (2,more hours)
        (5,other) };
  \end{axis}
\end{tikzpicture}
    \caption{Reasons cited for improved productivity when working from home, from 27 respondents.}
    \label{fig:home-good}
\end{figure}

Not unexpectedly, the distribution of factors that affect productivity was different for students and professionals.
Among students, the top factor that led to improved productivity was reduced stress.
Among professionals, by contradistinction, the top factor for improved productivity was by far reduced interruptions from co-workers.
In both groups the top factor that led to reduced productivity was lack of connections.

Not having to commute and working more hours, which were cited as factors that enhanced productivity, were write-ins.
It is possible that if we had suggested them in advance more participants would have ticked them off.

\section{Discussion}

Software development productivity depends on many factors.
One is the physical setup of the workstation where the development is done.
Another is the familiarity and comfort of the developer with this setup.
As a result it is hard to perform reliable controlled experiments where developers must work in different conditions:
Some of the setups will necessarily be unfamiliar or uncomfortable, causing a confounding effect.

We therefore conducted a subjective study about how developers perceive the influence of multiple monitors on coding productivity.
This was partly based on a natural experiment, where developers work in different setups in the course of their everyday work.
One part of this natural experiment involved differences between work and home setups.
Another involved changes made to home setups in response to the Covid-19 pandemic, and observations about what caused changes in productivity when working from home due to the pandemic.

Our survey results lead us to conclude that, subjectively, programmers tend to feel that multiple-monitor settings benefit their productivity.
It seems that the current state of the practice in the industry is to use two monitors per workstation.
At the same time, most people have only one monitor in their home workstation, but would like to have more than one.
We believe, based on our analysis, that the main reason for settling for a less comfortable setting is the more limited space that developers typically have at home, which makes using multiple monitors impractical (see Figs.\ \ref{fig:desk} and \ref{fig:why-diff}).

But the Covid-19 natural experiment showed that using multiple monitors, and more generally the physical setup of the workstation, is not the dominant factor.
We expected that distractions at home, such as caring for kids or the elderly, might be a problem.
But the main result was that the most important factor is co-workers: lack of interactions with co-workers was the main reason for both an increase and a decrease in productivity when working at home.

This points to the fact that workplaces offer more than facilities. 
Human interaction with other co-workers is also important.
But as our results indicate, the direction of the effect is inconsistent.
Increased productivity can occur if one relies on co-workers for professional discussions, advice, and guidance.
A decrease in productivity may happen if your work is interrupted too often by others, whether for social or professional reasons.
The trade-off between socially distracting and professionally beneficial interactions should be investigated in future work, among other questions regarding the optimal work environment for programmers.

\section{Threats to Validity}

\paragraph{Threats to construct validity}

A major threat to construct validity is obviously that we don't measure productivity directly.
Instead productivity was assessed subjectively by asking the participants to assess changes in their productivity under different circumstances.
This method of measurement is simple and intuitive, but may suffer from bias, because different participants have different standards and beliefs of what is considered ``productive''.
In fact, there is even a debate whether is should be quantified by the number of code lines, the level of functionality, or something else.

However, we believe our approach is still preferrable to the alternative of deciding on a metric and conducting a controlled experiment.
This has been attempted in the past (e.g.\ \cite{czerwinski03,colvin04,kang08}).
However, all these experiments suffer from drawbacks, such as using an artificial sequence of tasks.
This is problematic not only because these tasks may not be representative of what people naturally do, but also because they are externally motivated.
When performing real work, part of a developer's task is to decide what to do.
Following arbitrary instructions is quite different.

\paragraph{Threats to internal validity}

Just like we can't be sure that our participants subjective perceptions of productivity match their real productivity, we can't be sure that their perceptions of causality match the real factors that affect their productivity.
We observed several correlations between factors and performance, but correlation could be caused by a third factor.
For example, company size or culture could influence both productivity and the physical working conditions.

Moreover, a sample of 101 participants is not very large, and when they are partitioned into groups (e.g.\ those that use single or dual monitors in daily work) the sample becomes even smaller.
In small groups there is a larger danger that statistical fluctuations will look meaningful.

\paragraph{Threats to external validity}

Threats to the generalizability of our results are mainly due to the many possible biases in the answers received.
These biases result from the fact that we used personal ties to recruit participants in the survey.
This leads to a concentration of participants with similar ages, backgrounds, experience, and education.
In particular, many of our participants actually come from a small set of well established companies.
Thus they may be used to higher standards than in smaller companies.

\section{Conclusions}

Referring to a computer's monitor as a ``desktop'' is more wishful thinking than reality.
Desks provide \emph{much} more space than monitors.
Common tools such as IDEs and browsers use tabs in an attempt to compensate for the lack of space.
This allows developers to have multiple open files at once.
But it is not the same as actually being able to see all the files in parallel, as one could with papers on a physical desk (or dining table \cite{henderson86}).
The popularity of using multiple monitors reflects on this craving for real physical space.
Being closer to a real desktop reduces cognitive load (you don't have to remember what's behind a tab).
It also allows for the allocation of different places for different activities, instead of timesharing the same space.

Furthermore, both our work and previous research show that people tend to use a specific monitor more frequently than the others in their workstation.
Multiple monitors provide not only space but also an organization of space, with different applications and tasks assigned to distinct monitors.
But the way that this is done (or alternatively, the ways that applications are moved around) is personal, with different developers exhibiting different preferences and habits.
This implies that the effect on productivity may also depend on individual differences.

Using multiple monitors is now the standard in software engineering jobs.
This situation should be taken into account in software engineering research.
For example, it is desirable to develop cross-monitor eye tracking, to enable a deeper study of real work practices in realistic settings.

Furthermore, new tools are required that make good use of the additional space.
One practical idea is for IDEs to delegate some panels to secondary monitors (e.g.\ browsing the project directories, running tests), leaving the main monitor free for displaying more than one file at a time with large windows.
This would facilitate empirical studies of how such a design affects work habits, and in particular studies that can form the basis for more reliable results on the interplay between using multiple monitors and productivity.

\section*{Data Availability}

The survey form and results are available on the IEEE DataPort at URL http://dx.doi.org/10.21227/95tx-e080.

\bibliographystyle{myabbrv}
\bibliography{abbrv,se}

\end{document}